\documentclass[preprint2]{aastex}

\newcommand{\etal}{\rm et al.}
\newcommand{\pone}{Paper\,{\small{\sc I}}}
\newcommand{\ptwo}{Paper\,{\small{\sc II}}}

\newcommand{\hi}{H\,{\small{\sc I}}}

\usepackage{epsfig}
\usepackage{graphics}

\tighten

\lefthead{M. Dahlem et al.}
\righthead{Ambiguous fits to X-ray data of NGC\,253 and M\,82}

\begin{document}
 
\title{Ambiguities in fits to the complex X-ray spectra of starburst
  galaxies}

\author{Michael Dahlem, Arvind Parmar, Tim Oosterbroek, and Astrid Orr}
\affil{European Space Research and Technology Centre (ESTEC), Astrophysics 
  Division, \\ Space Science Department, Postbus 299, NL-2200 AG 
  Noordwijk, The Netherlands}
\and
\author{Kimberly A. Weaver}
\affil{Code 662 NASA/Goddard Space Flight Center, Greenbelt MD 20771}
\and
\author{Timothy M. Heckman}
\affil{Johns Hopkins University, Department of Physics and Astronomy,
Homewood Campus, \\ 3400 North Charles Street, Baltimore, MD 21218-2695}

\begin{abstract}

Spectral fits to X-ray data from both NGC\,253 and M\,82 provide
ambiguous results. The so-called ``best fit'' results depend on 
the instrument with which the data were obtained and obviously 
on the choice of spectral model composition. We show that different 
spectral models can be fit equally well to {\it BeppoSAX} data 
of both galaxies. Metallicities are unreliable in general, with 
a strong dependence on the choice of model. Preference to one 
particular spectral model can only be given by combining 
spectroscopic and imaging X-ray data from all available satellites 
({\it ROSAT}, {\it ASCA}, and {\it BeppoSAX}). Based on spectra 
of NGC\,253, we demonstrate that a model consisting of 
two or more thermal plasma components plus a hard power law 
continuum and Fe K$\alpha$ line emission can explain all 
observations. These model components represent the integral 
spectrum of thermal gas and compact sources in starburst 
galaxies that are most likely supernova remnants and X-ray 
binaries. The same model can fit the X-ray data of M\,82,
but there the evidence, from {\it ROSAT}\ imaging, of the
existence of compact sources which might represent high-mass
X-ray binaries is weaker. This implies that its hard X-ray
emission, which is extended in {\it ROSAT}\ images, might--if
truly diffuse--be dominated by a very hot (several keV energy)
thermal gas component.

\end{abstract}


\section{Introduction}
\label{par:intro}

There have been a number of papers recently discussing {\it ASCA}
and {\it ROSAT}\ X-ray observations of nearby starburst galaxies, 
in particular the two best-studied systems NGC\,253 and M\,82, with 
different interpretations of the resulting spectral fits (Moran \& 
Lehnert 1997; Ptak \etal\ 1997; Strickland \etal\ 1997; Tsuru \etal\ 
1997; Vogler \& Pietsch 1999). All authors agree on the general
complexity of the X-ray properties, but--depending on the data and 
on the spectral models used for the spectral analysis--they reached 
different conclusions. Especially the choice of spectral model 
components and the resulting best-fitting element abundances are 
under debate.

By treating all available {\it imaging and spectroscopy} data from 
{\it ASCA} and {\it ROSAT}\ (both HRI and PSPC) in a self-consistent 
manner, Dahlem \etal\ (1998; hereafter \pone) and Weaver \etal\ 
(2000; hereafter \ptwo) were able to reconcile the apparent 
discrepancies based on a mini-survey of 5 nearby edge-on starburst 
galaxies.

The results from \pone\ and \ptwo\ suggest that the combined {\it 
ASCA} and {\it ROSAT}\ PSPC integral spectra of NGC\,253 and M\,82 
can be fit with comparable values of $\chi^2$ by different 
combinations of spectral components, which means that there is an 
ambiguity in the choice of the best-fitting spectral model. By 
cross-checking the spectral results with {\it ROSAT}\ PSPC and HRI 
imaging data, a spectral composition of (at least) two thermal 
plasmas, with temperatures in the ranges 0.1--0.4 keV and 0.6--0.8 
keV, respectively, plus a hard power law component turns out to be 
the only model combination that can explain {\it all} observational 
data of all galaxies in the sample simultaneously (\ptwo).

This is in contrast with the recent findings by Cappi \etal\ (1999;
hereafter C99), based on {\it BeppoSAX} data of NGC\,253 and M\,82. 
These authors claim that there is ``compelling evidence for the 
presence of an extended hot thermal gas'' of several keV temperature 
in these two galaxies. The purpose of the current letter is to 
investigate this apparent discrepancy between their results and 
ours by re-analyzing the {\it BeppoSAX} observations of NGC\,253 
and M\,82, taking into account earlier results based on {\it ROSAT}\ 
and {\it ASCA} observations.

\section{Observations and data reduction} 

All parameters of the {\it BeppoSAX} observations of M\,82 and 
NGC\,253 are as described by C99. The data were reduced in the 
standard fashion, using {\it SAXDAS 2.0}. LECS and MECS data 
were extracted for joint spectral fitting from a circular region
centered on the position of the sources, using radii of $4'$. 
Background subtraction was performed using standard files (Parmar,
Oosterbroek, \& Orr 1999 ***check***).

Spectral fitting was performed using {\sc xspec} in the following 
way. We first used the input model preferred by C99 to ensure that 
we can reproduce their results. Then we tried the model used by us 
for the analysis of the joint {\it ASCA}+{\it ROSAT}\ PSPC spectrum 
(\pone\ and \ptwo). 

In the following we will list and discuss our results for both 
NGC\,253 and M\,82. However, since we used the same data 
extraction, reduction, and spectral fitting technique for both 
galaxies, only one (NGC\,253) will be presented in figures.

\section{Results and discussion}

\subsection{NGC\,253}

\begin{figure*}[t!]
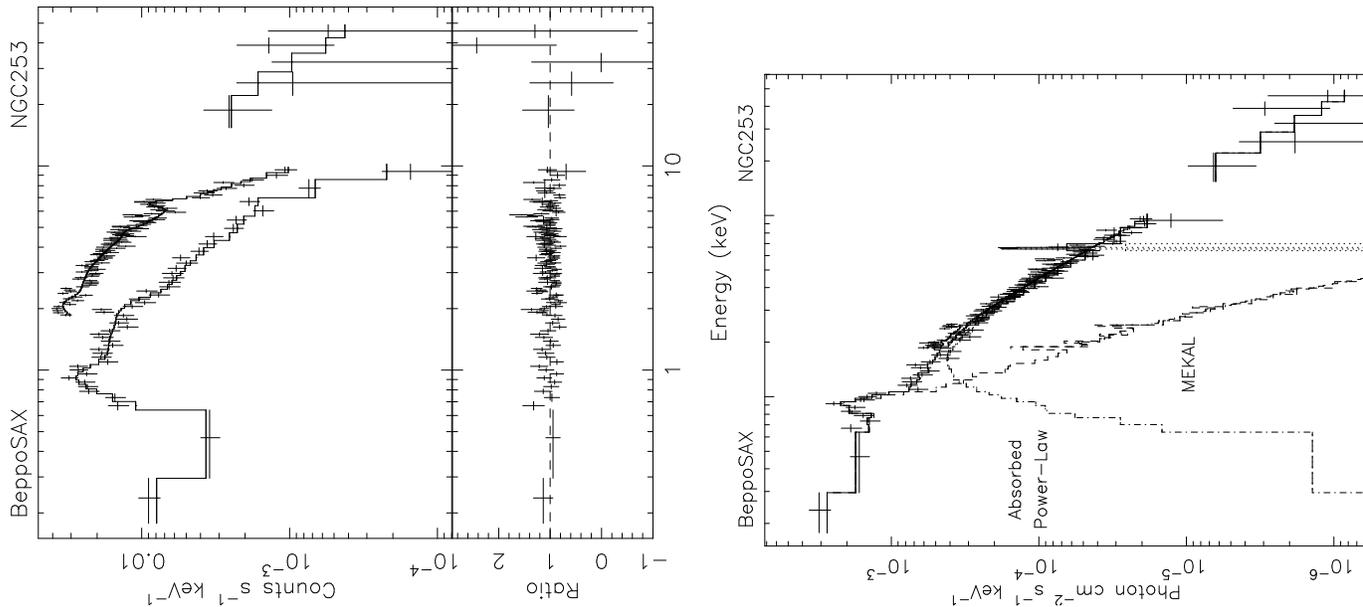

\hspace*{1.5cm}
\epsfig{file=n253_w2000fit.ps, angle=0, height=8cm}
\epsfig{file=n253_w2000unf.ps, angle=0, height=7.1cm}
\figcaption[ ]{
Fit of the model by Weaver \etal\ (2000) to the {\it BeppoSAX} 
observations of NGC\,253; the model fit to the data (left) and 
the unfolded spectrum (right) are displayed.
\label{fig:kimfit}}
\end{figure*}

We could reproduce the results by C99 within the uncertainties,
using a spectral model with 2 Mekal plasma components (2M; see 
their Fig.~4). The fit to the {\it BeppoSAX} data of NGC\,253 
following the model preferred by us, with a Mekal and a power 
law (M+P) component, is displayed in Fig.~\ref{fig:kimfit}. This 
is evidently also an acceptable fit. The goodness of fit for our 
preferred model (\ptwo) is $\chi^2$ = 261.9 for 265 d.o.f. 
($\chi^2/\nu$ = 0.99), while that for the 2 Mekal (hereafter
``2M'') model favored by C99 is $\chi^2$ 
= 282.5 for 268 degrees of freedom (d.o.f.) and thus $\chi^2/\nu$ 
= 1.05. The M+P model fits the data better than the 2M model at 
the highest and lowest energies of the passband. The results of 
the two spectral fits to the {\it BeppoSAX} data of NGC\,253 
are tabulated in Table~\ref{tab:fits}. All uncertainties are 
given at the 90\% confidence level for one interesting 
parameter; note that these only apply under the assumption
that the choice of model components represents the different
contributing emission mechanisms correctly.

The softest thermal emission component found in the {\it ROSAT}\ 
data is not required. When including a thermal plasma of 0.26 keV 
energy, $\chi^2$ is improved, but not significantly. Therefore 
it was left out in the fits to the {\it BeppoSAX} data. 

The hard part of the X-ray spectrum can be fit with a power 
law that is compatible with those of Galactic X-ray binaries
(XRBs) and can thus be explained naturally as the continuum 
emission from HMXRBs (\ptwo). The integral spectrum of all 
the point sources from the {\it ROSAT}\ PSPC observations 
is consistent with this interpretation, as is the contribution 
of this spectral component to the total X-ray flux (\pone). 
There is no reason why all {\it compact} sources should emit 
a thermal spectrum. 

The claim that there is a hot thermal gas at a few keV energy 
(C99) hinges only on the assumption that this is the only 
mechanism that could explain the observed Fe line around 6.7 
keV. However, supernova remnants (SNRs) and XRBs, including 
high-mass XRBs (HMXRBs), {\it can} produce both fluorescent 
and thermal Fe line emission, i.e., at 6.4 keV and 6.7 keV, 
respectively (e.g., Nagase 1989; White, Nagase, \& Parmar 1995; 
Liedahl \etal\ 1999), while C99 assume that all line emission 
is of thermal origin. Thus, the observed Fe line might well
be a superposition of emission from hot gas and from X-ray
binaries. The fitted equivalent width of the Fe line 
(Table~\ref{tab:fits}) is in agreement with this interpretation.
It, too, might be a composition of a (broad) thermal component
and a (narrow) XRB contribution. However, a composite line
fit cannot be performed based on the current data, because
the line is only marginally resolved.

The above finding that different models can fit the data equally 
well proves that the ``optimal'' fit is model-dependent, as 
already stated in \ptwo. Therefore, there is no reason to
reject the M+P model. Moreover, it is, as argued in \ptwo, the 
physically most plausible model choice.

Taking into account the {\it ROSAT}\ imaging results, which 
indicate clearly that there is a considerable number of 
unresolved compact sources in the central part of the disk 
of NGC\,253 (\pone), the most likely identification of these 
sources--based on their spectral properties and soft X-ray 
luminosities, $L_{\rm X}$--is that they constitute a population 
of HMXRBs (\pone). Thus, part of the emission distribution seen 
by {\it BeppoSAX} is not truly ``diffuse'', but smeared out 
by its broad point-spread function.

These point sources detected by {\it ROSAT}\ contribute about
50\% of the flux from the central disk (\pone\ and \ptwo). 
Thus, they are very significant contributors to the measured 
total flux, especially in the hard part of the X-ray spectrum.

On the other hand, the spectral model preferred by C99 does 
{\bf not} take into account the presence of HMXRBs and their 
spectral signature. Given the luminosities of emission 
mechanisms tracing the presence of high-mass stars in galaxies 
like NGC\,253 and M\,82, especially far-infrared radiation, a 
large number of HMXRBs must be expected to be present in them.

It is still unclear how the previously detected X-ray emitting 
thermal plasma (with temperatures in the range of a few tenths 
of a keV) is heated, especially in the galaxy halos, up to 
several kpc away from the disk planes of the starbursts. The 
presence of another, extremely hot medium of several keV energy 
contributing of order 2/3 of the total 2--10 keV flux, as 
suggested by C99, would further excruciate the problem of 
energy supply.

When taking into account the trade-off between metallicities
and absorbing \hi\ column densities in fitting the softest 
part of X-ray spectra (which cannot be resolved by {\it 
BeppoSAX} data only, but requires the low-energy response
of {\it ROSAT}), extreme subsolar metallicities, $Z$, are 
{\bf not} required to obtain a good fit (\ptwo). This $N_{\rm 
H}$ vs. $Z$ dichotomy is another, independent ambiguity in 
the minimum $\chi^2$ space of the spectral fits. Low 
metallicities in starburst galaxies, i.e., the galaxies with 
the highest star formation rates in the local Universe, would 
be hard to understand because of the proven presence of large 
numbers of massive stars, which are the most prolific producers 
of metals.

\subsection{M\,82}

The same ambiguities are present in fits to the {\it BeppoSAX}
data of M\,82. We could fit almost equally well the model by C99 
and ours from \pone\ and \ptwo. Just as for NGC\,253, the M+P 
model fits the data points at the very highest energies slightly
better than a 2M model. The goodness of fit is 517.4/446 d.o.f. 
= 1.16 (2M model) and 466.9/442 d.o.f. = 1.06 (M+P model), 
respectively. Note that, just as for the combined {\it ROSAT}\ 
+ {\it ASCA} data, the {\it BeppoSAX} data require another, soft 
thermal component to be added to the M+P model. 

With the M+P spectral model composition, we obtain almost equally 
good fits with two very different metallicities. In one case, 
$Z = 17 Z_\odot$ (constrained to be $>2.2$ at the 90\% confidence 
level), in the other $Z = 0.13 Z_\odot$. In the high-metallicity
case the flux at $\sim 1$ keV is modeled primarily as Ne and
Fe-L line emission, while in the low-metallicity case it is
modeled as a peak in the thermal distribution. The energy
resolution of the LECS of $\sim 200$ eV ({\it FWHM}) at 1 keV 
is insufficient to discriminate between the two options. Note 
that these two fits do not yet take into account the additional 
information obtained with {\it ASCA} and {\it ROSAT}\ requiring
an additional soft thermal component (\pone\ and \ptwo). 

There is less evidence from {\it ROSAT}\ imaging for the existence 
of large numbers of compact sources in M\,82. Instead, there appears 
to be a spatially extended, hard spectral component. Part of this 
might be truly diffuse, in which case the most likely interpretation
is that of a very hot gaseous component, as suggested by C99.
Only recently the {\it Chandra} image by Griffiths et al. (2000)
showed that there is indeed a population of compact sources in
M\,82, surrounded by diffuse emission. The compact sources in
the central part of M\,82 could not be resolved by {\it ROSAT},
because they are too close to each other. The most likely 
identification is again that they are HMXRBs (Griffiths et al.
2000). Individual HMXRBs could also explain the observed X-ray 
variability in the hard part of the spectrum, while there is no
evidence in the {\it Chandra} data for the presence of an AGN 
(Ptak \& Griffiths 1999, Matsumoto \& Tsuru 1999, Gruber \& 
Rephaeli 1999, C99).

The measured position and equivalent width of the Fe line in M\,82 
of $6.63\pm0.21$ keV and $60\pm40$ eV, respectively, leaves open
whether the line emission comes from either binaries or diffuse
hot gas or a superposition of both. In both M\,82 the width of the 
Fe line near 6.6 keV is unresolved. Thus, except 
for the (poorly constrained) position of the line centroid, no 
further information on the relative contribution of thermal or 
fluorescent line emission can be made based on the existing data. 
Both model compositions tested above fit the data (statistically) 
so well that no useful constraint can currently be derived on the 
possible contribution of both a hot thermal plasma and HMXRBs to 
the 2--10 keV flux of M\,82, when fitted simultaneously.

\section{Summary}

There are several ambiguities in the fits to complex X-ray 
spectra of starburst galaxies, such as NGC\,253 and M\,82. The 
``best-fitting'' model is not necessarily unique, because the 
spectral models required to explain all observations are more 
complex than can be fit unambiguously to one single dataset. 
In such cases statements that a fit is good at a certain
significance level can be misleading, because they only apply
if the correct spectral model composition was chosen.
There are also intrinsic degeneracies, i.e., trade-offs of
different fit parameters against each other (e.g., $N_{\rm H}$
vs. $Z$).

This study makes it clear how important it is to consider all
available information, including in particular X-ray imaging 
results of extended sources, when interpreting their integral 
spectral properties. The new generation of X-ray satellites, 
{\it Chandra}, {\it XMM}, and {\it Astro-E}, will resolve much 
of this ambiguity because of their high spectral resolution, 
combined with high sensitivity and good imaging capabilities 
over wide bandpasses, rendering possible spatially resolved 
spectroscopy of individual (classes of) sources within nearby 
galaxies.


\begin{deluxetable}{lccccccccc}
%
\tablewidth{0pt}
%
\tablecaption{Comparison of fit results to {\it BeppoSAX} data of NGC\,253
\label{tab:fits}}
\tablecolumns{9}
\tablehead{ 
\colhead{Source$^1$} & \colhead{Model$^2$} & \colhead{kT$_{\rm m}^3$} 
  & \colhead{N$_{\rm H,m}$} & \colhead{Z$_{\rm m}$} & \colhead{kT$_{\rm 
  h}$/$\Gamma$} & \colhead{N$_{\rm H,h}$} & \colhead{Z$_{\rm h}$} 
  & \colhead{$\chi^2/\nu$} \\
\colhead{} & \colhead{} & \colhead{(keV)} & 
  \colhead{($10^{22}$ cm$^{-2}$)} & \colhead{($Z_\odot$)} & 
  \colhead{(keV/...)} & \colhead{($10^{22}$ cm$^{-2}$)} & 
  \colhead{($Z_\odot$)} & \colhead{} 
} 
\startdata
C99   & 2M  & $0.65^{+0.09}_{-0.07}$ & $0.33^{+0.22}_{-0.07}$ & 
  $0.25^{+0.06}_{-0.09}$ & $5.90^{+0.9}_{-0.4}$ & $0.33^{+0.22}_{-0.07}$ & 
  $0.25^{+0.06}_{-0.09}$ & 1.05 \\
\ptwo & M+P & $0.75\pm0.06$ & 0.013(f) & $0.16\pm0.02$ 
  &$2.22\pm0.04^4$ & $1.11\pm0.11$ & \nodata & 0.99 \\
\tablenotetext{1}{C99 = Cappi et al. (1999); \ptwo\ = Weaver et al. (2000)}
\tablenotetext{2}{2M = Mekal + Mekal; M+P = Mekal + Power Law}
\tablenotetext{3}{Subscript m for ``medium'' energy thermal component
  (\ptwo), subscript h denotes the alleged ``hot'' thermal component
  (C99)}
\tablenotetext{4}{Fe K$\alpha$ line fitted independently at $6.62\pm0.07$
  keV, with an equivalent width of $EW = 480\pm110$ eV}
\tablenotetext{f}{Parameter fixed}
\enddata
\end{deluxetable}

%
%

%

\end{document}